\documentclass[12pt,letterpaper]{article}
\usepackage{latexsym}
\usepackage{amssymb}
\usepackage{amsmath}
\usepackage{slashed}

\evensidemargin=\oddsidemargin
\newcommand{\half}{\frac{1}{2}}
\newcommand{\be}{\begin{equation}}
\newcommand{\ee}{\end{equation}}
\newcommand{\bea}{\begin{eqnarray}}
\newcommand{\eea}{\end{eqnarray}}
\newcommand{\beat}{\begin{eqnarray*}}
\newcommand{\eeat}{\end{eqnarray*}}
\newcommand{\bi}{\begin{itemize}}
\newcommand{\ei}{\end{itemize}}
\newcommand{\bid}{\begin{displaymath}}
\newcommand{\eid}{\end{displaymath}}

\newcommand{\prm}{\textrm{ .}}
\newcommand{\crm}{\textrm{ ,}}

\newcommand{\Dp}{$\overline{\textrm{D$p$}}$ }

\newcommand{\ie}{{\em i.e.}}
\newcommand{\eg}{{\em e.g.}}
\newcommand{\refb}[1]{(\ref{#1})}
\newcommand{\imply}{\Rightarrow}

\def\half{\frac{1}{2}}

\def\eps{\epsilon}

\def\appendix{{\section*{Appendix}}\let\appendix\section%
        {\setcounter{section}{0}
        \gdef\thesection{\Alph{section}}}\section}

\catcode`\@=11
\def\@citex[#1]#2{%
\if@filesw \immediate \write \@auxout {\string \citation {#2}}\fi
\@tempcntb\m@ne \let\@h@ld\relax \def\@citea{}%
\@cite{%
  \@for \@citeb:=#2\do {%
    \@ifundefined {b@\@citeb}%
      {\@h@ld\@citea\@tempcntb\m@ne{\bf ?}%
      \@warning {Citation `\@citeb ' on page \thepage \space undefined}}%
      {\@tempcnta\@tempcntb \advance\@tempcnta\@ne%
      \@tempcntb\number\csname b@\@citeb \endcsname \relax%
      \ifnum\@tempcnta=\@tempcntb 
        \ifx\@h@ld\relax%
          \edef \@h@ld{\@citea\csname b@\@citeb\endcsname}%
        \else%
          \edef\@h@ld{\ifmmode{-}\else--\fi\csname b@\@citeb\endcsname}%
        \fi%
      \else
        \@h@ld\@citea\csname b@\@citeb \endcsname%
        \let\@h@ld\relax%
      \fi}%
    \def\@citea{,\penalty\@highpenalty\,}%
  }\@h@ld
}{#1}}

\def\@citeb#1#2{{[#1]\if@tempswa , #2\fi}}
\def\@citeu#1#2{{$^{#1}$\if@tempswa , #2\fi }}
\def\@citep#1#2{{#1\if@tempswa , #2\fi}}

\def\bcites{         
        \catcode`\@=11
        \let\@cite=\@citeb
        \catcode`\@=12
}

\def\upcites{         
        \catcode`\@=11
        \let\@cite=\@citeu
        \catcode`\@=12
}

\def\plaincites{      
        \catcode`\@=11
        \let\@cite=\@citep
        \catcode`\@=12
}

\begin{document}
\begin{titlepage}
\begin{flushright}
hep-th/0510205\\
HIP-2005-45/TH \\
\end{flushright}
\vspace*{3mm}
\begin{center}
{\Large {\bf On Superstring
Disk Amplitudes}\\}
{\Large \bf in a Rolling Tachyon Background\\}
\vspace*{12mm} {\bf Niko Jokela$^*$}\footnote{\tt e-mail: niko.jokela@helsinki.fi},
{\bf Esko Keski-Vakkuri$^{*,**}$}\footnote{\tt e-mail: esko.keski-vakkuri@helsinki.fi},
and {\bf Jaydeep Majumder$^*$}\footnote{\tt e-mail: jaydeep.majumder@helsinki.fi}\\
\vspace{3mm}
{\it $\mbox{}^*$Helsinki Institute of Physics and $\mbox{}^{**}$Department of Physical Sciences\\[.1cm]
P.O. Box 64, FIN-00014 University of Helsinki, Finland}
\vspace{1cm}

\begin{abstract}
 \noindent We study the tree level scattering or emission of $n$
closed superstrings from a decaying non-BPS brane in Type II superstring theory. We attempt to calculate
generic $n$-point superstring disk amplitudes in the rolling tachyon background. We show that these
can be written as infinite power series of Toeplitz determinants, related to expectation values of
a periodic function in Circular Unitary Ensembles. Further analytical progress is possible in the
special case of bulk-boundary disk amplitudes. These are interpreted as probability amplitudes for
emission of a closed string with initial conditions perturbed by the addition of an open string
vertex operator. This calculation has been performed previously in bosonic string theory, here
we extend the analysis for superstrings. We obtain a result for the average energy of closed
superstrings produced in the perturbed background.
\end{abstract}
\end{center}
\setcounter{footnote}{0}
\baselineskip16pt


\end{titlepage}

\section{Introduction}

One of the basic open questions in string theory is understanding
the decay of unstable branes. Sen has proposed a CFT description for
spatially homogenous decay by deforming the open string worldsheet
theory by exactly marginal rolling tachyon backgrounds
\cite{Sen:2002nu,Sen:2002in,Sen:2002an}. This process can be
interpreted as a spacelike brane localized in time (full
S-brane) \cite{Gutperle:2002ai}. An alternative, rescaled rolling
tachyon background \cite{Larsen:2002wc} corresponds to decay
starting from past infinity (half S-brane). One can also consider brane decay
on a space-time orbifold with a semi-infinite time
direction, to obtain a model where the unstable brane is prepared at origin of time
and then decays \cite{Kawai:2005jx}.
Basic questions such as
computing amplitudes for scattering or emission of strings from
decaying branes have turned out to lead into quite complicated
calculations rendering it difficult to draw out lessons of physics
interest. Several different approaches to this problem have been
explored, such as timelike boundary Liouville theory
\cite{Gutperle:2003xf} and matrix integrals \cite{Constable:2003rc}.
In particular, for full S-branes, a prescription based upon analytic continuation
to imaginary time where the full S-brane corresponds to an array of smeared
D-branes, was proposed in \cite{Gaiotto:2003rm}.
Further references include
\cite{Teschner:2001rv,Mukhopadhyay:2002en,Okuda:2002yd,Strominger:2002pc,Lambert:2003zr,Strominger:2003fn,Sen:2003bc,Schomerus:2003vv,Karczmarek:2003xm,Sarangi:2003sg,Okuyama:2003jk,Fredenhagen:2003ut,Hashimoto:2004zj,Hashimoto:2004qp,Fredenhagen:2004cj,Gaberdiel:2004na,Takayanagi:2004yr,Shelton:2004ij,Nakayama:2004ge,Kristjansson:2004ny,Lee:2005ge,Giribet:2005ix},
and the recent reviews \cite{Sen:2004nf,Schomerus:2005aq}.
Recently, for half S-branes,
this problem was elaborated and mapped into the study of random
matrices \cite{Balasubramanian:2004fz}. In this paper we extend this
approach to a study of superstring scattering from unstable branes
in superstring theory. It would be interesting to compare the random
matrix approach
with that of \cite{Gaiotto:2003rm}.

The general setup is also interesting from the point of view of
cosmology. Recently, there has been progress in constructing string
theoretic models of inflation. Of particular motivational interest
here are models based in Type IIB superstring theory, where inflation
arises from interactions of branes in (single or multiple) warped
throats
\cite{Kachru:2003sx},\cite{Barnaby:2004gg,Kofman:2005yz,Frey:2005jk,Chialva:2005zy}.
In these models, it has been proposed
\cite{Barnaby:2004gg,Kofman:2005yz,Frey:2005jk,Chialva:2005zy} that
reheating after inflation is associated with KK modes of gravitons
that are produced copiously as end decay products of massive closed
strings emitted from decaying $D\bar D$-systems at the throats.
However, the emission of massive closed strings is at present under
calculational control only for production of single strings, see
\cite{Lambert:2003zr,Karczmarek:2003xm,Sarangi:2003sg,Shelton:2004ij}.
In this
paper we aim for progress in calculating closed string $n$-point
disk amplitudes in the rolling tachyon background in superstring
theory, that could be interpreted as probability amplitudes for
multi-string emission. This is a very complicated problem, and we
are able to make only partial progress.

One technique to organize these calculations is to map them
to a computation in the language of random matrices: the amplitudes turn out to involve power
series of expectation values of periodic functions in Circular Unitary Ensembles (CUEs)
of $U(N)$ matrices
of increasing rank. This was found in \cite{Balasubramanian:2004fz}
in the context of two-point disk amplitudes in bosonic string theory; in this paper we generalize
the observation for generic $n$-point disk amplitudes in superstring theory.
Such expectation value calculations are a basic question in the theory of
random matrices. However, for the particular periodic functions that arise in the calculations,
the expectation values are only known\footnote{As far as we are aware of.} as Toeplitz determinants
of Fourier coefficients of the function. Further progress, needed for extracting physics lessons from the
amplitudes, is then associated with new progress in the field of random matrix theory
and mathematical analysis.

In the special case of bulk-boundary disk amplitudes, two-point functions of one bulk and one
boundary vertex operator, it is known that the calculations can be carried out to the point of
actually finding corrections to the one-point amplitude in an analytic form. These results have
been derived in bosonic string theory \cite{Balasubramanian:2004fz}, and also in
\cite{Schomerus:2003vv,Fredenhagen:2004cj} using Liouville theory methods. In this paper
we will extend the calculations and results to the case of superstrings.

This paper is organized as follows. In section 2, we consider generic $n$-point superstring disk
amplitudes and show how they are related to infinite power series of expectation values in
CUEs, or Toeplitz determinants of increasing rank. In section 3, we calculate the bulk-boundary
disk amplitudes in superstring theory. In section 4, we interpret the open string vertex operator
as an additional initial perturbation on the decaying brane, and calculate how it corrects the
average energy of the emitted closed strings in the decay. Finally, section 5 is a brief summary.

\section{Generic Closed String Disk Amplitudes and Random Matrices}

We begin by attempting to compute NS-NS and R-R disk amplitudes in the background
of a decaying brane. Depending on the external momentum assignments, these could
be interpreted as scattering or emission probability amplitudes.
As in \cite{Balasubramanian:2004fz,Shelton:2004ij}, we focus on the
$\half$S-brane or rolling tachyon background, which for the non-BPS brane
of Type II superstring corresponds to the exactly marginal deformation
\be\label{e1}
 \delta S_B = -\sqrt{2}\pi \lambda \int \frac{dt}{2\pi}\psi^0 e^{X^0/\sqrt{2}}\otimes\sigma_1 \ ,
\ee
where $\psi^0$ is the time component of the worldsheet
fermion field and $\sigma_1$ is a Chan-Paton factor associated with the boundary tachyon,
which can be related to the one-dimensional boundary
fermion $\eta$ \cite{Larsen:2002wc,Takayanagi:2001qh,Uesugi:2002vs,Shelton:2004ij}; see Appendix for
an elaboration on this point. For the bulk closed string vertex operators $V_s$, one can adopt
convenient gauge
choices \cite{Hwang:1991an,Lambert:2003zr} (see also \cite{deBoer:2003hd,Shelton:2004ij}), where the dependence on the time component $X^0$ of the bosonic field takes a simple form:
\be
  V_s = e^{i\omega_cX^0} V^{\perp}_s (X^i,\psi^i, \tilde{\psi}^i,\ldots)
\ee in the NS-NS sector, and
\be\label{e3}
  V_s = e^{i\omega_cX^0} \Theta_{s_0}\tilde{\Theta}_{\tilde s_0}
  V^{\perp}_s (X^i,\psi^i, \tilde{\psi}^i,\ldots)
\ee
in the R-R sector, with the spin fields $\Theta_{s_0}=e^{is_0H^0}$ in the bosonized form.
(The ellipsis refers to ghosts and superconformal ghosts.)
Thus, for a generic $n$-point closed string amplitude,
the non-trivial
part of the computation due to the presence of the rolling tachyon amounts to the expectation
value
\be\label{eq:AnNSNS}
 A_n(\omega_1,\ldots,\omega_n) \equiv \left\langle\prod_{a=1}^n e^{i\omega_a X^0(z_a,\bar z_a)}
 \right\rangle_{\rm{deformed}} =
\left\langle e^{\sqrt 2\pi\lambda\sigma_1\int_{-\pi}^\pi\frac{dt}{2\pi}\psi^0 e^{\frac{X^0}{\sqrt 2}}}
\prod_{a=1}^n e^{i\omega_a X^0(z_a,\bar z_a)} \right\rangle
\ee
for vertex operators in the NS-NS sector, and
\be\label{eq:AnRR}
 A_n(\omega_1,\ldots,\omega_n) \equiv \left\langle\prod_{a=1}^n e^{i\omega_a X^0(z_a,\bar z_a)}
 \Theta^{(a)}_{s_0} \tilde{\Theta}^{(a)}_{\tilde s_0}
 \right\rangle_{\rm{deformed}}
\ee
in the R-R sector. Consider for example the NS-NS sector disk amplitude in more detail.
Bosonizing the fermionic superpartner $\psi^0$ and expanding, we obtain (odd terms vanish, since ${\rm Tr}[\sigma_1^n] = 0$, for $n=\ odd$)
\be
 A_n(\omega_1,\ldots,\omega_n) = \int_{-\infty}^\infty dx^0 e^{ix^0\sum_{a=1}^n\omega_a}
 \sum_{N=0}^\infty\frac{\left(\pi\lambda e^{\frac{x^0}{\sqrt 2}}\right)^{2N}}{(2N)!}
{\cal O}(\omega_1,\ldots,\omega_n)\crm
\ee
where
\be\label{om}
{\cal O}(\omega_1,\ldots,\omega_n) = \int_{-\pi}^\pi \prod_{i=1}^{2N}\frac{dt_i}{2\pi}\left\langle\left[e^{iH(t_i)}-e^{-iH(t_i)}\right]
 e^{\frac{X'(t_i)}{\sqrt 2}}\prod_{a=1}^n e^{i\omega_a X'(z_a,\bar z_a)}\right\rangle\crm
\ee
and we have separated out the zero mode from the fluctuating part, $X^0 = x^0 + X'^0$,
and further dropped the superscript $0$. The Wick contractions in (\ref{om}) are easily calculated and
we obtain
\bea\label{omega}
{\cal O}(\omega_1,\ldots ,\omega_n) & = & \sum_{\{\epsilon_i\}=\pm} \int^{\infty}_{-\infty} dh\prod^{2N}_{i=1}
\left( \epsilon_i e^{i\epsilon_i h} \right)\int_{-\pi}^\pi\prod_{i=1}^{2N}\frac{dt_i}{2\pi}
\prod_{1\leq i<j\leq 2N}|e^{it_i}-e^{it_j}|^{1+\eps_i\eps_j}\nonumber\\
 &\cdot&\prod_{i=1}^{2N}\prod_{a=1}^n|1-z_a e^{-it_i}|^{i\sqrt 2\omega_a}\prod_{1\leq a<b\leq n}
 |z_a-z_b|^{-\omega_a\omega_b}\prod_{a,b=1}^n|1-z_a\bar z_b|^{-\frac{\omega_a\omega_b}{2}}
\eea
where we have separated out the zero mode $h$ from $H$. The integral over it enforces a constraint
\be
 \sum^{2N}_{i=1} \epsilon_i =0\prm
\ee
In the sum over ${\epsilon_i}=\pm$, all the combinations subject to the constraint
contribute equally to (\ref{omega}), as can be seen by an appropriate relabeling of the $t_i$'s. Thus,
we can choose
\be
  \epsilon_1,\ldots,\epsilon_N=+1 \ ; \ \epsilon_{N+1},\ldots ,\epsilon_{2N} = -1
\ee
and count the number of all equivalent terms. This is a random
walk problem, there are $(2N)!/(N!)^2$ such terms. The remaining integrals then factorize and we
can write
\be
{\cal O}(\omega_1,\ldots ,\omega_n) = (-1)^N(2N)!\prod_{1\leq a<b\leq n}|z_a-z_b|^{-\omega_a\omega_b}\prod_{a,b=1}^n|1-z_a\bar z_b|^{-\half\omega_a\omega_b}
I^2_{N}(\omega_1,\ldots,\omega_n)\crm
\ee
where $I_{N}$ is the integral
\be
 I_{N}(\omega_1,\ldots ,\omega_n) =
 \frac{1}{N!}\int_{-\pi}^\pi\prod_{i=1}^{N}\frac{dt_i}{2\pi}\prod_{1\leq i<j\leq N}
 |e^{it_i}-e^{it_j}|^{2}
 \prod_{i=1}^{N}\left[\prod_{a=1}^n|1-z_a e^{-it_i}|^{i\sqrt 2\omega_a}\right]\ .
\ee
The study of this type of integrals is a central question in the theory of random matrices \cite{Mehta}.
We can recognize it as the expectation value of a periodic function with respect to the Circular Unitary
Ensemble of $U(N)$ matrices,
\be \label{IN}
  I_{N} \equiv {\bf{E}}_{U(N)}\left\{ \prod^{N}_{i=1} f(t_i) \right\}\crm
\ee
where the periodic function is
\be\label{ft}
  f(t) = \prod_{a=1}^n |1-z_a e^{-it}|^{i\sqrt 2\omega_a}\prm
\ee
It contains the information about the locations (modular parameters) $z_a$ of the closed string
vertex operators and the on-shell energies $\omega_a$.
Alternatively, because of the factorization, we could have written the result
as a $U(N)\times U(N)$ integral as in \cite{Shelton:2004ij},
\be
 I^2_N = {\bf{E}}_{U(N)}\left\{\prod^{N}_{i=1}f(t_i)\right\}\cdot{\bf{E}}_{U(N)}
 \left\{\prod^{N}_{i=1}f(t_i)\right\}
 = {\bf{E}}_{U(N)\times U(N)}\left\{ \prod^{2N}_{i=1} f(t_i) \right\}\prm
\ee

The integrals (\ref{IN}) can then be evaluated by Heine's identity \cite{Balasubramanian:2004fz,Heine}
and rewritten as Toeplitz determinants of
the Fourier coefficients\footnote{For example, in the case of a 2-point function the Fourier
coefficients turn out to be related to Hypergeometric functions,
see \cite{Balasubramanian:2004fz}.} of $f$,
\be
 I_N= \det (\hat{f}_{(k-l)})_{1\leq k,l\leq N} \equiv D_{N}[\hat{f}]\crm
\ee
where
\be
 \hat{f}_{(k-l)} = \int \frac{dt}{2\pi} f(t)e^{i(k-l)t}\prm
\ee
Thus the amplitude becomes a Fourier transform of an infinite series of Toeplitz determinants,
\bea \label{ANS}
 A_n(\omega_1,\ldots,\omega_n) &=& \prod_{1\leq a<b\leq n}|z_a-z_b|^{-\omega_a\omega_b}
 \prod_{a,b=1}^n|1-z_a\bar z_b|^{-\half\omega_a\omega_b}\nonumber \\
 &\cdot& \int^\infty_{-\infty}dx^0 e^{ix^0\sum^n_{a=1}\omega_a} F(x^0;\omega_1,\ldots,\omega_n)
 \crm
\eea
where
\be \label{F}
 F(x^0;\omega_1,\ldots,\omega_n) = \sum^{\infty}_{N=0}
 (-\pi^2\lambda^2 e^{\sqrt 2 x^0})^N (D_{N}[\hat{f}])^2\prm
\ee
Unfortunately, the Toeplitz determinants are in general quite complicated so a more
detailed analysis of the infinite series is extremely difficult. For example, the radius of convergence
of (\ref{F}) is difficult to determine. By physics reasons, we expect the infinite series to converge
at least for sufficiently
early times (as then the amplitude approaches that for scattering from a
stable D-brane). It might also be possible to gain some further insight into the behavior of the series from
numerical methods. However, in order to perform the Fourier transform in (\ref{ANS}),
one would need an analytic expression for (\ref{F}) and then analytically continue beyond its expected
convergence radius, a much harder task.

For R-R sector the story is a bit modified. The amplitude (\ref{eq:AnRR}) becomes
\be\label{eq:AnRR2}
 A_n(\omega_1,\ldots,\omega_n) = \int_{-\infty}^\infty dx^0 e^{ix^0\sum_{a=1}^n\omega_a}
 \sum_{N=0}^\infty\frac{\left(\pi\lambda e^{\frac{x^0}{\sqrt 2}}\right)^N}{N!}
{\cal O}(\omega_1,\ldots ,\omega_n){\rm Tr}(\sigma_1)^{N+n}\crm
\ee
where
\be\label{RRoverlap}
 {\cal O}(\omega_1,\ldots,\omega_n) = \int_{-\pi}^\pi \prod_{i=1}^{N}\frac{dt_i}{2\pi}\left\langle\left[e^{iH(t_i)}-e^{-iH(t_i)}\right]
 e^{\frac{X'(t_i)}{\sqrt 2}}\prod_{a=1}^n e^{i\omega_a X'(z_a,\bar z_a)}
 e^{is_a H(z_a)}e^{i\tilde s_a\tilde H(\bar z_a)}\right\rangle\prm
\ee

In the R-R sector one has to explain why the amplitude with a single R-R vertex operator in the bulk
is non-vanishing for an odd number of insertion of boundary tachyon vertex
operators \cite{9808141,9809111,9904207}. From \refb{e1}, it is clear that the vertex operator
of the tachyon contains the Chan-Paton matrix $\sigma_1$.
The Chan-Paton Hilbert space is
two-dimensional, even for a {\em single} non-BPS D-brane,
since a non-BPS D$p$-brane of Type IIA(B) theory can be thought of as a bound
state of a  D$p$-\Dp-pair of Type IIB(A) theory.
For an odd number of tachyon vertex operator
insertions on the boundary, naively we expect that the amplitude vanishes because of the presence of the
factor ${\rm Tr}[\sigma^{2n+1}_1] = 0$, $n\ $= +ve integer. However, this is not the full story, at least
for bulk-boundary amplitudes involving R-R sector (which will be considered in section 3).

For concreteness, let us suppose we are considering a non-BPS D$p$-brane in Type IIA theory
(so $p$ is {\em odd}). It is obtained by taking a D$p$-\Dp-brane pair in Type IIB and modding it
out by $(-1)^{F_L}$, where $F_L$ is the left-moving spacetime fermion number. The R-R and R-NS sectors
of Type IIA can be thought of as `twisted sector' states under $(-1)^{F_L}$ orbifold in Type IIB theory.
For diagrams involving R-R operators it is easier if we stick to Type IIB orbifold  rather than Type IIA
language. The operator $(-1)^{F_L}$ does not act on the matter or ghost part of any open string vertex
operator, but it has an action on the $2\times2$ CP Hilbert space. Since under its action a BPS D$p$-brane
gets exchanged with a  \Dp-brane, the representation of $(-1)^{F_L}$ in the CP Hilbert space is $\sigma_1$.
So a Type IIA disk diagram  with some $N$ number of boundary tachyon vertex operators and a R-R vertex
operator inserted in the bulk, from Type IIB orbifold perspective, is equivalent to a disk diagram with a
cut, associated with the $(-1)^{F_L}$ operator, ending on the boundary. Due to above representation
of $(-1)^{F_L}$ in the CP Hilbert space, the trace part in the full amplitude gets another factor
of $\sigma_1$, where the cut hits the boundary, \ie, now the trace from the CP sector
is ${\rm Tr}[\sigma_1^{N+1}]$. This is non-vanishing only when $N\ =\ odd$. It is straightforward to
extend this procedure for $n$ insertions of R-R vertex operators in the bulk. The amplitude will then
be non-vanishing iff $(N\ +\ n) = even$. Thus, if $N$ and $n$ are {\em even} integers separately, the
amplitude is still non-vanishing.\footnote{The whole analysis can be done in terms of the GSO
operator $(-1)^F$ instead of $(-1)^{F_L}$, where $F$ is the left-moving worldsheet fermion number.
This is a bit involved; interested readers may consult ref. \cite{9809111}.}

For a correlation function in \refb{RRoverlap} with $n$ number of bulk R-R and $N$ of boundary tachyon
operator insertions, the zero mode integral from the temporal part imposes a constraint
\be
 \sum_{i=1}^N\eps_i = -\sum_{a=1}^n(s_0^{(a)} + \tilde s^{(a)}_0)\equiv k \in \mathbb Z\prm
\ee
By inspection one can see that $N$, $n$ and $k$ all have
the same (even or odd) parity.

Similar considerations as for the NS-NS amplitudes show that the
constraint can be satisfied in $N\choose \frac{N-k}{2}$ equivalent ways.
Omitting contractions which are not relevant for our discussion, the source-dependent part of the
amplitude again leads to a series of expectation values of periodic functions in CUE ensembles,
\be
 {\cal O}(\omega_1,\ldots,\omega_n) \sim
 {\bf{E}}_{U(\frac{N-k}{2})}\left\{ \prod^{(N-k)/2}_{i=1} f_-(t_i) \right\}\cdot
 {\bf{E}}_{U(\frac{N+k}{2})}\left\{ \prod^{(N+k)/2}_{i=1} f_+(t_i) \right\}\crm
\ee
where the periodic functions $f_{\mp}(t)$ resemble (\ref{ft})
but differ in their exponents. The underlying $U(\frac{N-k}{2})\times U(\frac{N+k}{2})$ structure was
found in \cite{Shelton:2004ij} in the case of generic 1-point amplitudes.
By Heine's identity, the source-dependent part of the amplitude
can again be rewritten as an infinite series of (products of) Toeplitz determinants of increasing rank.
This structure generalizes also to generic $(n_1+n_2)$-point disk amplitudes,
where $n_1$ ($n_2$) counts NS-NS (R-R) bulk insertions.

Further progress on disk amplitude calculations depends on new techniques, and we hope to return
to this problem in the future. However, it is known that
there are some special amplitudes, where an analytic solution can be found --
the bulk-boundary amplitudes. They were computed in the bosonic case in \cite{Balasubramanian:2004fz},
and we will next extend these results to superstrings.

\section{Bulk-Boundary Disk Amplitudes}

Let us consider the case $n=2$, and place\footnote{The vertex operator cannot be mapped
into the boundary by a conformal transformation.}
the other vertex operator into the boundary of the disk,
thus renaming $\omega_1\equiv \omega_c$, $\omega_2\equiv \omega_o$. We consider the
operator in the boundary to represent an additional open string. In other words, we will consider
the amplitudes
\be
A_{NSNS,NS}(\omega_c,\omega_o) \equiv \left\langle e^{i\omega_c X^0(z,\bar z)}
e^{i\omega_o X^0(t)}\right\rangle_{\rm{deformed}}
\ee
and
\be
A_{RR,NS}(\omega_c,\omega_o) \equiv \left\langle e^{i\omega_c X^0(z,\bar z)}
\Theta_{s_0} \tilde{\Theta}_{\tilde s_0}
e^{i\omega_o X^0(t)}\right\rangle_{\rm{deformed}} \ .
\ee
We can choose the bulk vertex operator to be inserted at the origin of the disk, $z=\bar z=0$, while
the location $t$ of the boundary vertex operator remains a free modular parameter to be integrated
over in the end.

\subsection{NS-NS Bulk Vertex Operator}

Consider first the case with a NS-NS bulk vertex operator.
The amplitude becomes
\be
 A_2(\omega_c,\omega_o) = \int_{-\infty}^\infty dx^0 e^{ix^0(\omega_o+\omega_c)}
 \sum_{N=0}^\infty(-\pi^2\lambda^2 e^{\sqrt 2 x^0})^N [I_N(\omega_o)]^2 \ ,
\ee
where $I_N(\omega_o)$ is the integral
\be \label{Iselberg}
 I_N(\omega_o) = \frac{1}{N!}\int_{-\pi}^\pi\prod_{i=1}^{N}\frac{dt_i}{2\pi}\prod_{1\leq i<j\leq N}|e^{it_i}-e^{it_j}|^2
 \prod_{i=1}^{N}|1-e^{it}e^{-it_i}|^{i\sqrt 2\omega_o}\prm
\ee

We have removed an apparent divergence resulting from the self-contractions on the boundary,
by an appropriate normal ordering \cite{Larsen:2002wc}.
The multiple integrals over $t_i$ do not depend on $t$, hence we can
set $t=0$. As noted in \cite{Balasubramanian:2004fz}, the integral $I_N$ can be evaluated using
Selberg's integral formula. After some algebra, we can then evaluate the amplitude in a closed form
in terms of known functions. Defining a ``chemical potential''
$\mu =-\log(-\pi^2\lambda^2 e^{\sqrt 2 x^0})$, carefully following the
calculational strategy in \cite{Balasubramanian:2004fz}, and carrying out the $x^0$ integral using the
real contour of \cite{Lambert:2003zr} we obtain
\bea
 A_2(\omega_c,\omega_o) & = & \int_{-\infty}^\infty dx^0 e^{ix^0(\omega_o+\omega_c)}\sum_{N=0}^\infty e^{-N\mu}\left[\prod_{j=1}^{N}\frac{\Gamma(j)\Gamma(j+i\sqrt 2\omega_o)}{(\Gamma(j+i\omega_o/\sqrt 2))^2}\right]^2 \nonumber\\
 & = & \int_{-\infty}^\infty dx^0 e^{ix^0(\omega_o+\omega_c)}\sum_{N=0}^\infty
 e^{-N\mu} e^{2\int_0^\infty dt H(t,\omega_o/\sqrt 2)(e^{-Nt}-1)} \nonumber\\
 & = & \frac{-i\pi}{\sqrt 2}
 \frac{(\pi\lambda)^{-i\sqrt 2(\omega_o+\omega_c)}}{\sinh(\pi (\omega_o+\omega_c)/\sqrt 2)}
 \exp \left\{2\cdot G\left(\frac{\omega_c}{\sqrt 2},\frac{\omega_o}{\sqrt 2}\right)\right\}\crm
\eea
where
\be
 G\left(\frac{\omega_c}{\sqrt 2},\frac{\omega_o}{\sqrt 2}\right)
\equiv \int_0^\infty dt H\left(t,\frac{\omega_o}{\sqrt 2}\right)(e^{i(\omega_o+\omega_c)t/\sqrt 2}-1)\crm
\ee
with
\be
H(t,\omega_o) \equiv\frac{(1-e^{-i\omega_o t})^2}{2t(1-\cosh t)}\crm
\ee
similar to the result in \cite{Balasubramanian:2004fz}. As a simple consistency check we can verify
that the result reduces\footnote{Apart from an irrelevant
overall phase factor $-i$.} to the answer in \cite{Shelton:2004ij}
in the absence of the initial open string perturbation, $\omega_o=0$. This follows easily
since $H(t,\omega_o)$ vanishes in the limit.

\subsection{R-R Bulk Vertex Operator}
Consider then the R-R closed string vertex operator \cite{Shelton:2004ij}
\be
 \Theta_{s_0}\tilde\Theta_{\tilde s_0}e^{i\omega_c X^0(z_c,\bar z_c)}\crm
\ee
and bosonize the spin fields
\be
 \Theta_{s_0} = e^{is_0 H^0} \ ; \ \tilde{\Theta}_{\tilde s_0} = e^{i\tilde s_0\tilde H^0} \prm
\ee
Note that in the series expansion of the amplitude the terms with $N=even$
vanish. The relevant Wick contractions now give
\bea
 & &\left\langle \prod_i(e^{iH(t_i)}-e^{-iH(t_i)})e^{is_0H(0)}e^{i\tilde s_0\tilde H(0)} e^{i\omega_c X'(0,0)}e^{i\omega_o X'(t)}e^{X'(t_i)/\sqrt 2}\right\rangle \nonumber\\
 & = & -2s_0\delta_{s_0,\tilde s_0}\frac{(-1)^N(2N+1)!}{N!(N+1)!}\left[\prod_{1\leq i<j\leq N}|e^{it_i}-e^{it_j}|^2\right]\left[\prod_{N+1\leq i<j\leq 2N+1}|e^{it_i}-e^{it_j}|^2\right]\nonumber\\
&\cdot& \left[\prod_{i}|1-e^{it}e^{-it_i}|^{i\sqrt 2\omega_o}\right]
 \times \rm{(irrelevant\ terms)}\prm
\eea
Note that we have already integrated out the zero modes of $H$ and $\tilde H$. We have suppressed the
details of terms that will ultimately not contribute because the bulk vertex operator has been placed at
the origin.
The amplitude becomes
\bea
 A_2(\omega_c,\omega_o) = -2s_0~\delta_{s_0,\tilde s_0}\int^\infty_{-\infty}dx^0e^{ix^0(\omega_o+\omega_c)}
 \sum^{\infty}_{N=0}(-1)^N(\pi\lambda e^{x^0/\sqrt 2})^{2N+1} I_{N}\cdot I_{N+1} \crm
\eea
where $I_N,I_{N+1}$ are the same Selberg integrals as before, giving
\be
 I_N = \prod^N_{j=1} \frac{\Gamma (j)\Gamma (j+i\sqrt 2 \omega_o)}{(\Gamma (j+i\omega_o/\sqrt 2))^2} \ .
\ee
Proceeding as before, we get
\bea
 A_2(\omega_c,\omega_o) &=& -2s_0~\delta_{s_0,\tilde s_0}\int^\infty_{-\infty}dx^0e^{ix^0(\omega_o+\omega_c)}
 \sum^{\infty}_{N=0}(-1)^N(\pi\lambda e^{x^0/\sqrt 2})^{2N+1} \nonumber \\
 &\cdot &
 \exp\left\{\int^\infty_0 dt~H(t,\omega_o/\sqrt 2)[e^{-Nt}(1+e^{-t})-2)]\right\} \crm
\eea
and, after some algebra, finally
\bea\label{RRdisk}
 A_2(\omega_c,\omega_o) &=&
 -\frac{2s_0\pi\delta_{s_0,\tilde s_0}}{\sqrt 2}\frac{(\pi\lambda)^{-i\sqrt 2(\omega_o+\omega_c)}}
 {\cosh(\pi(\omega_o+\omega_c)/\sqrt 2)} \nonumber \\
  && \cdot \exp \left\{G\left(\frac{\omega_c}{\sqrt 2}-\frac{i}{2},\frac{\omega_o}{\sqrt 2}\right)+
    G\left(\frac{\omega_c}{\sqrt 2}+\frac{i}{2},\frac{\omega_o}{\sqrt 2}\right)\right\}\prm
\eea
Note that this again meets the result in \cite{Shelton:2004ij} as $\omega_o\to 0$.

We would like to add a few comments on the delta function present in
the equation \refb{RRdisk}. It implies that the left- and right-movers in
the R-R field are such that $s_0 = \tilde{s}_0$. It results from the
correlation function of the spin field along the temporal direction
of the R-R vertex operator, $V_s$, given in equation \refb{e3}.
Apparently, it does not contain any information about the nature of
the theory, \ie, whether this result holds in either Type IIA or
Type IIB or both. Certainly, such information can not come from the
temporal part of the correlation function. These informations are
contained in other parts of the vertex operators, which we have suppressed since they do not
take part in the physics of the rolling tachyon. First, the full R-R
vertex operator has a spatial part, $V^{\perp}_s$, as defined in
\refb{e3}, which depends on the spin fields along spatial
directions. Second, the R-R vertex operator also has a piece with R-R
field strength given by
\be\label{RRfield}
\slashed{F}_{\alpha\beta} \sim
F_{\mu_1\cdots\mu_k}(\Gamma^{\mu_1}\cdots\Gamma^{\mu_k})_{\alpha\beta}\crm
\ee
where $\alpha$, $\beta$ are spinor indices ($\alpha,\beta =
1,\ldots,32$) and we suppressed the normalization constants which
are not so important for our purpose. Finally, there is another
piece which also contributes Gamma matrices. This comes from the
standard doubling trick procedure for such bulk-boundary correlation function computations,
where we extend the definition of the holomorphic field from upper half-plane (UHP)
to lower half-plane (LHP) by equating it to its anti-holomorphic partner:
\be\label{DT}
X(z) = \left\{
\begin{array}{lcl}
X(z)\,, \ \ \qquad\text{for}\quad z\in\text{UHP}\,,\\
\pm\widetilde{X}(z)\,,\qquad\text{for}\quad z\in\text{LHP}
\end{array}
\right.
\ee
where the $\pm$ sign is for Neumann (Dirichlet) directions.
For a spin field, it gives
\be\label{Doubling}
\widetilde{S}^{\alpha}(\bar{z}) =
(\Gamma^0\cdots\Gamma^p)^{\alpha}_{\ \beta}\, S^{\beta}(\bar{z})\crm
\ee
where $p$ is the dimensionality of the D$p$-brane, and for a non-BPS
brane $p = odd\ (even)$ for Type IIA (IIB). Once we take all this into
consideration, the restriction on the sets $\{s_0,s_i\}$ and
$\{\tilde{s}_0,\tilde{s}_i\}$ turns out to be\footnote{For our purpose we choose the
chirality of left- and right-moving R sector spinors in such a way
that for Type IIA : $\sum_i s_i =even$ and $\sum_i \tilde{s}_i =
odd$, whereas for Type IIB it is : $\sum_i s_i = \sum_i \tilde{s}_i
= even$.}
\bea\label{quantumno}
\text{For Type IIA}\ :\qquad s_0 + \tilde{s}_0     & = & 0 \imply s_0 = -\tilde{s}_0\nonumber\\
 \quad\qquad\sum_{i=1}^4( s_i + \tilde{s}_i)       & = & \pm 1 \\
\!\!\text{For Type IIB}\ :\qquad s_0 + \tilde{s}_0 & = & \pm 1 \imply s_0 = \tilde{s}_0 = \pm \half\nonumber\\
 \sum_{i=1}^4( s_i + \tilde{s}_i)                  & = & 0\crm
\eea
so that the Type IIA (IIB) spinor chiralities can be satisfied correctly.

\section{Energy Emission}
We are ultimately interested in computing the expectation value of total emitted energy from the decaying brane.
For the unperturbed initial state of the D-brane (spatially homogeneous decay), it was found in \cite{Shelton:2004ij} that the total energy of closed strings emitted was divergent for a D$p$-brane with $p\leq 2$. We now shortly examine how this is modified when the initial state is perturbed by addition of the boundary tachyon vertex operator, thus extending the discussion in \cite{Balasubramanian:2004fz} to superstring.

So the relevant question is, how does the inclusion of an open string perturbation change the asymptotics of brane decay into closed strings? For this we need the asymptotics of $G(\omega_c,\omega_o)$ for $\omega_c\gg \omega_o$. Using a method
described in \cite{KeatingSnaith} we find for large $n$ that
\be
 e^{G(in+is/2,-is/2)} = \prod_{j=1}^n\frac{\Gamma(j)\Gamma(j+s)}{(\Gamma(j+s/2))^2}\sim n^{(s/2)^2}e^{(s/2)^2(\gamma+1)+\sum_{j=3}^\infty (-s)^j\frac{2^{j-1}-1}{2^{j-1}}\zeta(j-1)}\prm
\ee
So upon analytic continuation\footnote{Assuming that the ratio $e^{G(in+is/2,-is/2)} / n^{(s/2)^2}$ is analytic around $n=\infty$.} we get the asymptotics with large $\omega_c\gg \omega_o$,
\be
 2G\left(\frac{\omega_c}{\sqrt 2},\frac{\omega_o}{\sqrt 2}\right) \sim -2\omega_o^2\log\left(\frac{\omega_c}{\omega_o}\right)
\ee
for NS-NS bulk amplitude, and
\be
 G\left(\frac{\omega_c}{\sqrt 2}-\frac{i}{2},\frac{\omega_o}{\sqrt 2}\right)+
    G\left(\frac{\omega_c}{\sqrt 2}+\frac{i}{2},\frac{\omega_o}{\sqrt 2}\right) \sim -2\omega_o^2\log\left(\frac{|\omega_c+i/\sqrt 2|}{\omega_o}\right) \sim -2\omega_o^2\log\left(\frac{\omega_c}{\omega_o}\right)
\ee
for R-R bulk amplitude. The total emitted energy is calculated by
summing over all emitted closed string energies \cite{Lambert:2003zr,Balasubramanian:2004fz}
\be
 \frac{\overline E}{V_p} = \sum_s \half |A_2(\omega_c,\omega_o)|^2 \sim \frac{1}{(2\pi)^p}\int d\omega_c \omega_c^{-p/2-2\omega_o^2}\crm
\ee
showing that the result is in close analogy to bosonic case.

The lesson is that, without the perturbation $\omega_o=0$, emitted energy diverges for $p<3$ and
is finite for $p\geq 3$.
But morally speaking, we expect divergencies to indicate that the unstable brane decays
completely into closed strings, whereas for finite emitted energy into a lower dimensional brane.
The extra factor $\omega_c^{-2\omega_o^2}$ is a suppression (enhancement)
if $\omega_o$ is real (imaginary), depending on the dimension of the D$p$-brane. However, for
perturbations with imaginary $\omega_o$, decay into closed strings is enhanced, so we expect a
complete decay to closed strings for all $p$. See \cite{Balasubramanian:2004fz} for additional discussion.

\section{Summary}

We have investigated superstring disk amplitudes in the rolling tachyon background corresponding
to an eternally decaying non-BPS brane. Such computations address very basic questions about
how these branes decay.
We have shown here that the general structure of the amplitudes
is a Fourier transform of a power series in the target space time
coordinate, where the coefficients are Toeplitz determinants arising from expectation values
of a periodic function in Circular Unitary Ensembles of increasing rank. The periodic function
encodes the essential information about the amplitude. The determinants of increasing
rank $N$ compute disk amplitudes with $N$ open string tachyon vertex operators from the rolling tachyon
background. Further progress is related to advance in solving mathematical problems
in the context of random matrices, in particular
there is a need to investigate grand canonical ensembles, where the rank of the ensemble
(corresponding to the number of open string tachyon insertions) can
vary. So far, the calculations can be carried out fully only in the special case of bulk-boundary
amplitudes.
Apart from the more difficult mathematical problems, one tractable direction
to pursue could be to study the field theory limit of the power series of the individual terms,
and compare it with results computed from the effective Dirac-Born-Infeld field theory, in
the spirit of earlier such investigations
(see, \eg, \cite{Kutasov:2003er,Niarchos:2004rw,Coletti:2004ri,Laidlaw:2004cf}).

\bigskip

\noindent {\bf Acknowledgments: } We are grateful to Vijay Balasubramanian, Per Kraus and
Asad Naqvi for discussions.
N.J. was supported in part by Magnus Ehrnrooth foundation.
E.K-V. was supported in part by the Academy of Finland.

\renewcommand{\theequation}{\thesection.\arabic{equation}}
\@addtoreset{equation}{section}
\addcontentsline{toc}{section}{Appendix}
\appendix{Relations between boundary fermions and Pauli matrices $\sigma_i$}

Recall that $\eta,\bar\eta$ are in fact Grassmann variables:
\bea
 \{\eta,\bar\eta\}   & = & 1 \\
 \eta^2 = \bar\eta^2 & = & 0\prm
\eea
Their spinorial representation on the two-dimensional Hilbert space can be easily worked out,
and is given by
\bea
\eta & = & \left(\begin{array}{cc} 0 & 1 \\ 0 & 0 \end{array}\right)\,,\qquad
 \bar\eta =  \left(\begin{array}{cc} 0 & 0 \\ 1 & 0 \end{array}\right) \crm\nonumber\\
\eta\bar\eta & = & \left(\begin{array}{cc} 1 & 0 \\ 0 & 0 \end{array}\right)\,,\qquad
 \bar\eta\eta =  \left(\begin{array}{cc} 0 & 0 \\ 0 & 1 \end{array}\right) \prm
\eea
Defining $\sigma_\pm = \half(\sigma_1\pm i\sigma_2)$, we find
\bea
\sigma_+ & = & \eta\,,\qquad \sigma_-  =  \bar\eta\,,\qquad
\sigma_3  = [\eta,\bar\eta]\nonumber\\
\sigma_1 & = &  \eta + \bar\eta\,,\qquad
 \sigma_2 =  -i(\eta-\bar\eta) \prm
\eea
The relevant part of the supersymmetric boundary action in terms of $\eta$, $\bar\eta$ on
a brane-anti-brane pair is
\be
\delta S_B \sim i \sqrt{\frac{2}{\pi}}\int_{\partial\Sigma} dt\
\big(\bar\eta\psi^{\mu}D_{\mu}T - \psi^{\mu}\eta\overline{D_{\mu}T}\big)(t) \prm
\ee
On a brane-anti-brane pair, the tachyon $T$ is a complex field. Substituting $T = U + iV$ in the above, we get
\be
\delta S_B \sim i \sqrt{\frac{2}{\pi}}\ \int_{\partial\Sigma} dt\
\big[(\bar\eta - \eta)\psi^{\mu}D_{\mu}U + i(\bar\eta + \eta)\psi^{\mu}D_{\mu}V\big] \prm
\ee
Next, to obtain a non-BPS D-brane from a brane-anti-brane pair, we choose the $(-1)^{F_L}$ projection
in such a way that only the 2nd term in the above equation
gets projected in. Choosing $V \sim \sqrt{2\pi}\lambda e^{X^0/\sqrt 2}$, we arrive at \refb{e1}.
The boundary fermion $\eta$ used in \cite{Shelton:2004ij} is actually $(\eta + \bar\eta) = \sigma_1$
in our notation.

\end{document}